\documentclass[aps,showpacs,preprintnumbers,amsmath,amssymb]{revtex4}

\usepackage{graphics,epsfig}
\usepackage{graphicx}
\usepackage{dcolumn}
\usepackage{bm}

\begin{document}

\title{What kinds of coordinate can keep the Hawking temperature
invariant for the static spherically symmetric black hole?}
\author{Chikun Ding and Jiliang  Jing}
\thanks{Corresponding author, Electronic address:
jljing@hunnu.edu.cn}
\affiliation{ Institute of Physics and
Department of Physics,
Hunan Normal University, Changsha, Hunan 410081, P. R. China \\
and
\\ Key Laboratory of Low Dimensional Quantum Structures and
Quantum Control of Ministry of Education, Hunan Normal University
, Changsha, Hunan 410081, P.R. China}

 \baselineskip=0.65 cm

\vspace*{0.2cm}
\begin{abstract}
\vspace*{0.2cm} By studying the Hawking radiation of the most
general static spherically symmetric black hole arising from
scalar and Dirac particles tunnelling, we find the Hawking
temperature is invariant in the general coordinate representation
(\ref{arbitrary1}), which satisfies two conditions: a) its radial
coordinate transformation is regular at the event horizon; and b)
there is a time-like Killing vector.

\end{abstract}

 \vspace*{1.5cm}

\maketitle

\section{introduction}

In recent years, a semi-classical method of modeling Hawking
radiation as a tunnelling effect has been developed and has excited
a lot of interest
\cite{man,man1,wil,ag,sh1,sh2,sh3,mn,vagenas,par,par1,par2,
pkra,pkra1,pkra2,arz,qqj,zz,sqw,ajm,pm1,pm}. Tunnelling provides not
only a useful verification of thermodynamic properties of black
holes but also an alternate conceptual means for understanding the
underlying physical process of black hole radiation. In the
tunnelling approach, the particles are allowed to follow classically
forbidden trajectories, by starting just behind the horizon onward
to infinity. The particles must then travel necessarily back in
time, since the horizon is locally to the future of the static or
stationary external region. The classical, one-particle action
becomes complex, signaling the classical impossibility of the
motion, and gives the amplitude an imaginary part, provides a
semi-classical approximation to free field propagators. In general
the tunnelling methods involve calculating the imaginary part of the
action $I$ for the (classically forbidden) process of s-wave
emission across the horizon, which in turn is related to the
Boltzmann factor for emission at the Hawking temperature, i.e.
\begin{eqnarray}\Gamma\propto e^{-2\text{Im}I}=
e^{-E/T_H},\end{eqnarray}where $T_H$ is the Hawking temperature of
the black hole, $E$ is the energy of the tunnelling particles.

There are two different approaches that are used to calculate the
imaginary part of the action for the emitted particle. The first
method developed was the Null Geodesic Method used by Parikh and
Wilczek \cite{wil}; The other approach is the Hamilton-Jacobi
Ansatz used by Agheben {\it et al} \cite{ag} which is an extension
of the complex path analysis of Padmanabhan {\it et al}
\cite{sh1,sh2,sh3}. For the Hamilton-Jacobi ansatz it is assumed
that the action of the emitted scalar particle satisfies the
relativistic Hamilton-Jacobi equation. From the symmetries of the
metric one picks an appropriate ansatz for the form of the action
and plugs it into the relativistic Hamilton-Jacobi equation to
solve.

Since a black hole has a well defined temperature it should
radiate all types of particles like a black body at that
temperature. The emission spectrum therefore contains particles of
all spins such as Dirac particles. In this paper, we will use the
Hamilton-Jacobi ansatz method to calculate the Hawking
temperature.

Can the Hawking temperature keep invariant under any coordinate
transformation? At the first glance, the Hawking temperature is
invariant. However, this invariance has been lost in the following
isotropic coordinate \cite{ag,mn} for the Schwarzschild black hole
\begin{eqnarray}\label{isotropic} t\rightarrow t,~~r\rightarrow
\rho,~~\ln \rho=\int \frac{dr}{r\sqrt{1-\frac{2M}{r}}}.
   \end{eqnarray}And so the line element of the Schwarzschild black
hole becomes\begin{eqnarray}
ds^2=-\left(\frac{2\rho-M}{2\rho+M}\right)^2dt^2+\left(\frac{2\rho+M}{2\rho}\right)^4d\rho^2
+\frac{(2\rho+M)^4}{16\rho^2}d\Omega^2,
   \end{eqnarray}
and the horizon $\rho_H=M/2.$ Substituting it and
$\phi=e^{i[-Et+W(\rho)+J(\theta,\varphi)]/\hbar}$ into
Klein-Gordon equation\begin{eqnarray}\label{kg}
\frac{1}{\sqrt{-g}}\partial_\mu(\sqrt{-g}g^{\mu\nu}
\partial\nu\phi)
-\frac{m^2}{\hbar^2}\phi=0,
   \end{eqnarray}we can obtain\begin{eqnarray}
 \text{Im}W_{\pm}(\rho)&=&\pm\text{Im}\left[ \int
\frac{(2\rho+M)^3d\rho}{4\rho^2(2\rho-M)}\sqrt{E^2-(\frac{2\rho-M}
{2\rho+M})^2(m^2+g^{ij}J_iJ_j)}\right]\nonumber\\
&=&\pm4\pi ME.
   \end{eqnarray}The probability is\cite{sh1,sh2,sh3}\begin{eqnarray}
\Gamma=\frac{\Gamma_{out}}{\Gamma_{in}}\propto
\exp\Big[-4\text{Im}W_+\Big] =\exp\Big[-16\pi
ME\Big]=\exp\Big[-\frac{E}{T_H}\Big],
   \end{eqnarray}
since $W_-=-W_+$. Then the black hole's temperature is \cite{mn}
\begin{eqnarray}T_H=\frac{1}{16\pi M},\end{eqnarray}
   which is one-half of the standard Hawking temperature $T_H=1/8\pi
   M$. The example tell us that the invariance is missing in the isotropic
coordinate! The reason for the phenomenon comes from the
coordinate transformation (\ref{isotropic}) itself. In the radial
coordinate transformation
\begin{eqnarray}\ln \rho=\int \frac{dr}{r\sqrt{1-\frac{2M}{r}}}=\int
F(r)dr,\end{eqnarray}
 the function $F(r)=\frac{1}{r\sqrt{1-2M/r}}$
has singularity at the horizon $r=2M$. So it needs to discuss that
in which coordinates can Hawking temperature be invariant.

The purpose of this manuscript is to investigate the invariance of
the Hawking temperature of the most general static spherically
symmetric black hole from scalar and Dirac particles tunnelling in a
general coordinate representations. In order to do that, we
introduce the metrics of the static spherically symmetric black in
the two coordinates: Schwarzschild-like and a general coordinates.
This general coordinate should satisfy two conditions: a) its radial
coordinate transformation is regular at the event horizon; b) there
exists a time-like Killing vector.

The paper is organized as follows. In Sec. 2 the different
coordinate representations for the general static spherically
symmetric black hole are presented. In Sec. 3 the Hawking
temperature of the general static spherically symmetric black hole
for scalar particles tunnelling is investigated. In Sec. 4 the
Hawking temperature of the general static spherically symmetric
black hole from Dirac particles tunnelling is studied. The last
section is devoted to a summary.

\section{Coordinate representations for general static spherically symmetric black hole}

In this section we introduce two kinds of the coordinate
representations for the general static spherically symmetric black
hole, i. e. the Schwarzschild-like and  a general coordinates.

\subsection{Schwarzschild-like  coordinate representation}

In Schwarzschild-like coordinate the line element for the most
general static spherically symmetric black hole in four dimensional
spacetime is described by
\begin{eqnarray}\label{ghs}
ds^2=-f(r)dt_s^2+\frac{1}{g(r)}dr^2+R(r)(d\theta^2+\sin^2\theta
d\varphi^2),
   \end{eqnarray}
where $f(r)$, $g(r)$ and $R(r)$ are functions of $r$, and $t_s$ is
the Schwarzschild-like time coordinate.

Because the spacetime (\ref{ghs}) is a static and spherically
symmetric one, a time-like Killing vector field
$\xi^\mu=(1,0,0,0)$ exists. An interesting feature of the black
hole worthy of note is that the norm of the Killing field
$\xi^\mu$ is zero on the event horizon $r_H$ since the horizon is
a null surface and the vector $\xi^\mu$ is normal to the horizon.
Then, for the non-extreme case we have $f(r)=f_1(r)(r-r_H)$ and
$g(r)=g_1(r)(r-r_H)$, where $f_1(r)$ and $g_1(r)$ are regular
functions in the region $r_H<r<\infty$ and their values are
nonzero on the outermost event horizon.

\subsection{General coordinate representation}

In order to insure that there is a time-like Killing vector in the
spacetime, the most general coordinate $(v,~u,~\theta,~\varphi)$
that transform from the Schwarzschild-like coordinate (\ref{ghs})
is
 \begin{eqnarray}\label{arbitrary}
v=\lambda t_s+\int dr G(r),~~~u=\int dr F(r),
\end{eqnarray}
where $v$ is the time coordinate,  $u$ is the radial one, and the
angular coordinates remain unchanged; $\lambda$ is an arbitrary
nonzero constant which re-scales the time; $G$ is arbitrary
functions of $r$ and  $F$ is a regular function of $r$. The line
element (\ref{ghs}) in the new coordinate becomes
\begin{eqnarray}\label{arbitrary1}
&&ds^2=-\frac{f(r(u))}{\lambda^2}dv^2+2\frac{f(r(u))G(r(u))}
{\lambda^2F(r(u))}dudv\nonumber\\&&\qquad
+\frac{\lambda^2-f(r(u))g(r(u))G^2(r(u))}{\lambda^2g(r(u))F^2(r(u))}du^2
   +R(r(u))(d\theta^2+\sin^2\theta
  d\varphi^2).
\end{eqnarray}
We now show that two well-known coordinates, the Painlev\'{e} and
Lemaitre coordinates, are the spacial cases of the metric
(\ref{arbitrary1}).

\subsubsection{Painlev\'{e} coordinate representation} In the
transformation (\ref{arbitrary}), one sets
  $\lambda=1$, $G(r)=\sqrt{\frac{1-g(r)}{f(r)g(r)}}$ and
$F(r)=1$, the line element (\ref{arbitrary1}) becomes
  the Painlev\'{e} coordinate representation\cite{man1,ding}
\begin{eqnarray}\label{pan}
  \label{ds2}ds^2=-f(r)dt^2+2\sqrt{\frac{f(r)(1-g(r))}{g(r)}}dtdr+dr^2
  +R(r)(d\theta^2+\sin^2\theta
  d\varphi^2),
\end{eqnarray}where $t$ is the Panlev\'{e} time.
 The metric (\ref{pan}) has no singularity at g(r) = 0, so the
metric is regular at the horizon of the black hole. That is to say,
the coordinate complies with perspective of a free-falling observer,
who is expected to experience nothing out of the ordinary upon
passing through the event horizon.

\subsubsection{Lemaitre coordinate representation} In the
transformation (\ref{arbitrary}), one sets $\lambda=1$,
$G(r)=\frac{1}{2}\sqrt{\frac{g(r)}{f(r)(1-g(r))}}
+\sqrt{\frac{1-g(r)}{f(r)g(r)}}$ and
$F(r)=\frac{1}{2}\sqrt{\frac{g(r)}{f(r)(1-g(r))}}$, the line
element (\ref{arbitrary1}) at present becomes the Lemaitre
coordinate representation\cite{ding,ding2}
\begin{eqnarray}\label{Lm}
ds^2=-f(r)\big[dV^2+dU^2\big]+2\frac{f(r) (2-g(r))}{g(r)}dVdU
  +R(r)(d\theta^2+\sin^2\theta
  d\varphi^2),
\end{eqnarray}where $U$ is
Lemaitre radial coordinate and $V$ is the Lemaitre time one. We can
see that the Lemaitre coordinate is time-dependant system, suggests
that there could be a genuine particle production.

\vspace*{0.4cm}

\section{Temperature of general static spherically
symmetric black hole from scalar particles tunnelling}

We now investigate scalar particles tunnelling of general static
spherically symmetric black hole.

\subsection{Scalar particles tunnelling in Schwarzschild-like  coordinate}

Applying the WKB approximation
\begin{eqnarray}\label{ans}
\phi(t,r,\theta,\varphi)=\exp\Big[\frac{i}{\hbar}I(t,r,\theta,\varphi)+I_1(t,r,\theta,\varphi)
+\mathcal{O}(\hbar)\Big],
   \end{eqnarray}
to the Klein-Gordon equation (\ref{kg}),
 then, to leading order in $\hbar$ we get the
   following relativistic Hamilton-Jacobi equation
   \begin{eqnarray}\label{hj}
g^{\mu\nu}(\partial_\mu I\partial_\nu I)+m^2=0.
   \end{eqnarray}
As usual, due to the symmetries of the metric (\ref{ghs}) and
neglecting the effects of the self-gravitation of the particles,
there exists a solution in the form
   \begin{eqnarray}\label{ansatz}
I=-Et_s+W(r)+J(\theta,\varphi).
   \end{eqnarray}
Inserting Eq. (\ref{ansatz}) and the metric (\ref{ghs}) into the
Hamilton-Jacobi equation (\ref{hj}), we find
   \begin{eqnarray}\label{ww}
 W_{\pm}(r)&=&\pm \int
\frac{dr}{\sqrt{f(r)g(r)}}\sqrt{E^2-f(r)(m^2+g^{ij}J_iJ_j)},
   \end{eqnarray}
where $J_i=\partial_i I$, $i=\theta,\varphi$. One solution of the
Eq. (\ref{ww}) corresponds to the scalar particles moving away
from the black hole (i.e. ``+" outgoing) and the other solution
corresponds to particles moving toward the black hole (i.e. ``-"
incoming). Imaginary parts of the action can only come due the
pole at the horizon. The probability of a particle tunnelling from
inside to outside the horizon is \cite{sh1,sh2,sh3}
   \begin{eqnarray}\label{gamma}
\Gamma=\frac{\Gamma_{out}}{\Gamma_{in}}\propto
\exp\Big[-4\text{Im}W_+\Big] =\exp\Big[-\frac{E}{T_H}\Big],
   \end{eqnarray}
since $W_-=-W_+$. Integrating around the pole at the horizon leads
to
\begin{eqnarray}\label{radial}
\text{Im} W_+=\frac{\pi E}{\sqrt{f'(r_H)g'(r_H)}}.
   \end{eqnarray}
Substituting
   (\ref{radial}) into (\ref{gamma}), we obtain the Hawking temperature\begin{eqnarray}\label{HT}
T_H=\frac{\sqrt{f'(r_H)g'(r_H)}}{4\pi },
   \end{eqnarray}
which shows that the temperature of general static spherically
symmetric black hole is the same as previous works \cite{man,man1}.

 \subsection{Scalar particles tunnelling in general coordinate}\label{General}

Here we study the scalar tunnelling in a general  coordinate
(\ref{arbitrary1}). Employing the ansatz
 \begin{eqnarray}\label{action}
   I=-Ev+W(u)+J(\theta,\varphi)\end{eqnarray}
and substituting the metric (\ref{arbitrary1}) into the
Hamilton-Jacobi equation (\ref{hj}), we obtain
\begin{eqnarray}
  &&\left[g(r(u))G^2(r(u)) -\frac{\lambda^2}{f(r(u))}\right]E^2
 - 2g(r(u))G(r(u))F(r(u))E W'(u)\nonumber\\&&
 +g(r(u))F^2(r(u))\big[W'(u)\big]^2+g^{ij}J_iJ_j+m^2=0.\end{eqnarray}
Then $W'(u)$ is
\begin{eqnarray}\label{w(u)}
   W'_\pm(u)=\frac{G(r(u))}{F(r(u))}E\pm\frac{\sqrt{\lambda^2E^2-f(r(u))[g^{ij}J_iJ_j+m^2]}}
   {F(r(u))\sqrt{f(r(u))g(r(u))}}.\end{eqnarray}
We will study the temperature for two cases: $G(r(u))$ is regular
function and $G(r(u))$ has a pole at horizon.

\label{nopole}\subsubsection{$G(r(u))$ is regular function at
horizon}

When $G(r(u))$ is regular at horizon, $g_{uu}$ of metric
(\ref{arbitrary1}) shows that there is still a coordinate
singularity at the horizon $r_H$. From equation (\ref{w(u)}) we
get
\begin{eqnarray}
   \text{Im}W_\pm(u)&=&\text{Im}\int du\left\{\frac{G(r(u))}{F(r(u))}E
   \pm\frac{\sqrt{\lambda^2E^2-f(r(u))[g^{ij}J_iJ_j+m^2]}}
   {F(r(u))\sqrt{f(r(u))g(r(u))}}\right\}\nonumber\\&=&\pm \frac{\lambda E\pi}
   {\sqrt{f'(r_H)g'(r_H)}}.\end{eqnarray}We can see the
   Im$W_\pm(u)$ are like those in Schwarzschild-like coordinate. Using
\begin{eqnarray}
\Gamma\propto\exp[-4\text{Im}W_+]=\exp\left[-\frac{\lambda
E}{T_H}\right],
   \end{eqnarray}
   we can recover Hawking temperature (\ref{HT}).

\label{pole}\subsubsection{$G(r(u))$ has a pole at horizon}

When $G(r(u))$ has a pole at horizon, without loss of generality, it
can be expressed as  $G(r(u))=\frac{C(r(u))}
{\sqrt{f(r(u))g(r(u))}}+D(r(u))$, where $C(r(u))$ and $D(r(u))$ are
the regular functions at horizon. From Eq. (\ref{w(u)}), we
obtain\begin{eqnarray}\label{imwu}
   \text{Im}W_\pm(u)&=&\text{Im}\int du\left\{
   \frac{D(r(u))}{F(r(u))}E+\frac{C(r(u))E\pm\sqrt{\lambda^2E^2-f(r(u))[g^{ij}J_iJ_j+m^2]}}
   {F(r(u))\sqrt{f(r(u))g(r(u))}}\right\}\nonumber\\&=& \frac{(\frac{C(r_H)}{\lambda}\pm1)\lambda E\pi}
   {\sqrt{f'(r_H)g'(r_H)}}.\end{eqnarray}

In the following, we will consider two cases, i. e., $C(r_H)\neq
\lambda$ and $C(r_H)=\lambda$:

i) If $C(r_H)\neq \lambda$,  after substituting
$G(r(u))=\frac{C(r(u))}{\sqrt{f(r(u))g(r(u))}}+D(r(u))$ into
$g_{uu}$ of metric (\ref{arbitrary1}), it is easy to see that
there is still a coordinate singularity at horizon $r_H$, and the
probabilities are
   \begin{eqnarray}
\Gamma_{out}\propto\exp\left[-2
\frac{(\frac{C(r_H)}{\lambda}+1)\pi}
   {\sqrt{f'(r_H)g'(r_H)}}\lambda E\right],~~~\Gamma_{in}\propto\exp\left[-2 \frac{(\frac{C(r_H)}{\lambda}-1)\pi}
   {\sqrt{f'(r_H)g'(r_H)}}\lambda E\right].
   \end{eqnarray}
It is interesting to note that $\Gamma_{out},\Gamma_{in}$ are
different from that in the Schwarzschild-like coordinate, but the
total probability is
   \begin{eqnarray}
\Gamma=\frac{\Gamma_{out}}{\Gamma_{in}}\propto\exp\left[-
\frac{4\pi}
   {\sqrt{f'(r_H)g'(r_H)}}\lambda E\right],
   \end{eqnarray}
and the Hawking temperature (\ref{HT}) is also recovered.

ii) If $C(r_H)=\lambda$, we can write
$C(r(u))=\lambda+H(r(u))\sqrt{f(r(u))g(r(u))}$, where $H(r(u))$ is
a regular function at horizon. Then we have
$G(r(u))=\frac{\lambda}{\sqrt{f(r(u))g(r(u))}}+H(r(u))+D(r(u))$.
Substituting it into $g_{uu}$ of metric (\ref{arbitrary1}), we
find that there is no coordinate singularity at horizon $r_H$ now.
From Eq. (\ref{imwu}), we obtain
 \begin{eqnarray}
\text{Im}W_{+}(u)=\frac{2\pi }{\sqrt{f'(r_H)g'(r_H)}}\lambda
E,~~~~\text{Im}W_{-}(u)=0,
   \end{eqnarray}this implies that
  $\Gamma_{in}=1$. So the overall tunnelling
probability is
\begin{eqnarray}
   \Gamma=\Gamma_{out}\propto\exp[-2\text{Im} W_+]=\exp\left[-\frac{4\pi E}
   {\sqrt{f'(r_H)g'(r_H)}}\right].\end{eqnarray}
It is obviously that the Hawking temperature (\ref{HT}) is
recovered.

From above discussions we know that the  Hawking temperature of
general static spherically symmetric black hole arising from the
scalar particles tunnelling is invariant in the general coordinate
(\ref{arbitrary1}).

\section{Temperature of general static spherically
symmetric black hole from Dirac
particles tunnelling }

In this section, we study the Dirac particles tunnelling of the
black hole in the coordinates (\ref{ghs}) and (\ref{arbitrary1}).

\subsection{ Dirac particles tunnelling in
Schwarzschild-like coordinate}

For a general background spacetime, the Dirac equation is \cite{rmp}
\begin{eqnarray}\label{dirac}
&&\left[\gamma^\alpha
e^\mu_\alpha(\partial_\mu+\Gamma_\mu)+\frac{m}{\hbar}\right]
\psi=0,
 \end{eqnarray}
  with
  \begin{eqnarray}
&&\Gamma_\mu=\frac{1}{8}[\gamma^a,\gamma^b]e^\nu_ae_{b\nu;\mu},
  \nonumber  \end{eqnarray}
where $\gamma^a$ are the Dirac matrices and  $e^\mu_a$ is the
   inverse tetrad defined by
     $\{e_a^\mu\gamma^a,~~~e_b^\nu\gamma^b\}=2g^{\mu\nu}
\times1$. For the general static spherically symmetric  black hole
in the Schwarzschild-like metric (\ref{ghs}) the tetrad can be
     taken as
   \begin{eqnarray}\label{tetrad}
   e_a^\mu=diag\left(\frac{1}{\sqrt{f(r)}},\sqrt{g(r)},\frac{1}
   {\sqrt{R(r)}},\frac{1}{\sqrt{R(r)}\sin\theta}\right).
   \end{eqnarray}
  We employ the following ansatz for the Dirac field
   \begin{eqnarray}\label{psi}
  &&\psi_\uparrow=\bigg(\begin{array}{ccc}A(t_s,r,\theta,\varphi)\xi_\uparrow\nonumber\\
   B(t_s,r,\theta,\varphi)\xi_\uparrow\end{array}\bigg)
   \exp\big(\frac{i}{\hbar}I_\uparrow(t_s,r,\theta,\varphi)\big)
   =\left(\begin{array}{ccc}A(t_s,r,\theta,\varphi)\nonumber\\ 0\nonumber\\
   B(t_s,r,\theta,\varphi)\nonumber\\0\end{array}\right)
   \exp\big(\frac{i}{\hbar}I_\uparrow(t_s,r,\theta,\varphi)\big),\nonumber\\
   &&\psi_\downarrow=\bigg(\begin{array}{ccc}C(t_s,r,\theta,\varphi)\xi_\downarrow\nonumber\\
   D(t_s,r,\theta,\varphi)\xi_\downarrow\end{array}\bigg)
   \exp\big(\frac{i}{\hbar}I_\downarrow(t_s,r,\theta,\varphi)\big)
   =\left(\begin{array}{ccc}0\nonumber\\ C(t_s,r,\theta,\varphi)\nonumber\\
   0\\D(t_s,r,\theta,\varphi)\nonumber\end{array}\right)
   \exp\big(\frac{i}{\hbar}I_\downarrow(t_s,r,\theta,\varphi)\big),\nonumber\\
   \end{eqnarray}
where ``$\uparrow$" and ``$\downarrow$" represent the spin up and
spin down cases, and $\xi_{\uparrow}$ and $\xi_{\downarrow}$ are
the eigenvectors of $\sigma^3$. Inserting Eqs. (\ref{tetrad}),
(\ref{psi}) into the Dirac equation (\ref{dirac}) and employing
\begin{eqnarray}\label{ans3}
    I_\uparrow=-Et_s+W(r)+J(\theta,\varphi),
   \end{eqnarray}
to the lowest order in $\hbar$  we obtain
\begin{eqnarray}\label{aa}
  && -\frac{A}{\sqrt{f(r)}}E+\sqrt{g(r)}B W'(r)
   +mA=0,
   \\ \label{bb}
  && \frac{B}{\sqrt{R(r)}}(J_\theta+\frac{i}{\sin\theta}J_\varphi
   )=0,
   \\ \label{cc}
&& \frac{B}{\sqrt{f(r)}}E-\sqrt{g(r)}A W'(r)
   +mB=0,
  \\ \label{dd}
  &&  -\frac{A}{\sqrt{R(r)}}(J_\theta +\frac{i}{\sin\theta}J_\varphi
   )=0,
   \end{eqnarray}
where we consider only the positive frequency contributions
without loss of generality. Eqs. (\ref{bb}) and (\ref{dd}) both
yield ($J_\theta +\frac{i}{\sin\theta}J_\varphi$) = 0 regardless
of $A$ or $B$, implying that $J(\theta,\varphi)$ must be a complex
function. We therefore can ignore $J$ from this point (or else
pick the trivial $J = 0$ solution).

Consider first the massless case $m=0$, Eqs. (\ref{aa}) and
(\ref{cc}) give
\begin{eqnarray}
W_{\pm}(r)=\pm\int\frac{Edr}{\sqrt{f(r)g(r)}}.
 \end{eqnarray}
We therefore recover the expected Hawking temperature (\ref{HT})
in the massless case.

   In the massive case $m\neq0$,  Eqs. (\ref{aa}) and
   (\ref{cc}) show
   \begin{eqnarray}
   (\frac{A}{B})^2=\frac{\frac{E}{\sqrt{f(r)}}+m}
   {\frac{E}{\sqrt{f(r)}}-m}
   \end{eqnarray}
and
\begin{eqnarray}
   W_\pm(r)=\int\frac{E}{\sqrt{f(r)g(r)}}\frac{2(\frac{A}{B})}{1+(\frac{A}{B})^2}dr.
   \end{eqnarray}
Noting $\lim _{r\rightarrow r_H}(\frac{A}{B})^2=1$, we find
 that the result of integrating
around the pole for $W$ in the massive case is the same as the
massless case and we recover the Hawking temperature (\ref{HT}).

For the spin-down case the calculation is very similar to the
spin-up case discussed above. Other than some changes of sign, the
equations are of the same form as the spin up case. For both the
massive and massless spin down cases the Hawking temperature
(\ref{HT}) is obtained, implying that both spin up and spin down
particles are emitted at the same temperature.

\subsection{ Dirac
particles tunnelling in general coordinate}

We take
 \begin{eqnarray}\label{psi4}
  &&\psi_\uparrow=\bigg(\begin{array}{ccc}A(v,u,\theta,\varphi)\xi_\uparrow\nonumber\\
   B(v,u,\theta,\varphi)\xi_\uparrow\end{array}\bigg)
   \exp\big(\frac{i}{\hbar}I_\uparrow(v,u,\theta,\varphi)\big)
   =\left(\begin{array}{ccc}A(v,u,\theta,\varphi)\nonumber\\ 0\nonumber\\
   B(v,u,\theta,\varphi)\nonumber\\0\end{array}\right)
   \exp\big(\frac{i}{\hbar}I_\uparrow(v,u,\theta,\varphi)\big),\nonumber\\
   &&\psi_\downarrow=\bigg(\begin{array}{ccc}C(v,u,\theta,\varphi)\xi_\downarrow\nonumber\\
   D(v,r,\theta,\varphi)\xi_\downarrow\end{array}\bigg)
   \exp\big(\frac{i}{\hbar}I_\downarrow(v,u,\theta,\varphi)\big)
   =\left(\begin{array}{ccc}0\nonumber\\ C(v,u,\theta,\varphi)\nonumber\\
   0\\D(v,u,\theta,\varphi)\nonumber\end{array}\right)
   \exp\big(\frac{i}{\hbar}I_\downarrow(v,u,\theta,\varphi)\big)\nonumber\\
   \end{eqnarray}
where $   I_\uparrow=-Ev +W(u)+J(\theta,\varphi).$ For the line
element (\ref{arbitrary1}), we chose the tetrad\begin{equation}
\!\!\!\!\!\!\!\! e_a^\mu=\left(
\begin{array}{cccc}
\frac{\lambda}{\sqrt{f(r(u))}}  & \sqrt{g(r(u))}G(r(u))  & 0& 0 \\
0  &  \sqrt{g(r(u))}F(r(u)) &0   &0 \\
0  & 0  & \frac{1}{\sqrt{R(r(u))}}    &0 \\
0 & 0  & 0 & \frac{1}{\sqrt{R(r(u))}\sin\theta}
\end{array}
\right).
\end{equation}
Then the Dirac equation (\ref{dirac}) can be expressed as
\begin{eqnarray}
\Big[-\frac{\lambda}{\sqrt{f(r(u))}}A-\sqrt{g(r(u))}G(r(u))B\Big]E
   +\sqrt{g(r(u))}F(r(u))B W'(u)
   +mA=0,
   \\
\Big[\frac{\lambda}{\sqrt{f(r(u))}}B+\sqrt{g(r(u))}G(r(u))A\Big]E
   -\sqrt{g(r(u))}F(r(u))A B W'(u)
   +mB=0.
   \end{eqnarray}

For the case  $m= 0$, we find
   \begin{eqnarray}
W'(u)=\left[\frac{G(r(u))}{F(r(u))}E\pm\frac{\lambda
E}{\sqrt{f(r(u))g(r(u))}F(r(u))}\right],
   \end{eqnarray}
which is similar to Eq. (\ref{w(u)}). Taking the same method used
in the section \ref{General}, it is easy to get the Hawking
temperature (\ref{HT}).

For the case $m\neq 0$, we find
\begin{eqnarray}
(\frac{A}{B})^2=\frac{\frac{E}{\sqrt{f(r(u))}}+m}
{\frac{E}{\sqrt{f(r(u))}}-m}
   \end{eqnarray}
and $\lim _{u\rightarrow u_H}\frac{A}{B}=\pm1$. We have
   \begin{eqnarray}
W'(u)=\left[\frac{G(r(u))}{F(r(u))}E\pm\frac{2\lambda|\frac{A}{B}|
E}{\sqrt{f(r(u))g(r(u))}F(r(u))(\frac{A^2}{B^2}+1)}\right],
   \end{eqnarray}
which is similar to Eq. (\ref{w(u)}). Taking the same method used in
the section \ref{General}, we also find the same Hawking temperature
(\ref{HT}).

From above discussions we know that the  Hawking temperature of
general static spherically symmetric black hole arising from the
Dirac particles tunnelling is also invariant in the general
coordinate (\ref{arbitrary1}).

\section{summary}

The Hawking temperature of the Schwarzschild black hole in the
isotropic coordinate shows us that the temperature is not invariant.
What kinds of coordinate can keep the Hawking temperature invariant
for the general static spherically symmetric black hole? By studying
the Hawking radiation of the most general static spherically
symmetric black hole arising from scalar and Dirac particles
tunnelling, we find that it is invariant in the general coordinate
representation (\ref{arbitrary1}), which satisfies two conditions:
a) its radial coordinate transformation is regular at the event
horizon; and b) there is a time-like Killing vector.

We also find some other interesting results: 1) For the coordinate
representations which do not exist coordinate singularity, such as
the general coordinate (\ref{arbitrary1}) with $C(r_H)=\lambda$
(include the Painlev\'{e} (\ref{pan}) and Lemaitre (\ref{Lm})),
$W_+$ has a pole at the event horizon but $W_-$ has a well defined
limit at the horizon. Then the imaginary part of $W_-$ is zero due
to the imaginary parts of the action can only come from the pole and
the probability of a particle tunnelling from inside to outside the
horizon is described by $\Gamma= \Gamma_{out}$. 2) The mass of the
particles and the angular quantum number do not affect the Hawking
temperature for both scalar and Dirac particles. 3) When time
coordinate transforms from $t_s$ to $\lambda t_s$, i. e., we
re-scale the time, the corresponding energy $E$ of the total
tunnelling particles is increased by $\lambda$ times, so re-scale of
the time does not affect the Hawking temperature.

\begin{acknowledgments}This work was supported by the
National Natural Science Foundation of China under Grant No.
10675045; the FANEDD under Grant No. 200317; the Hunan Provincial
Natural Science Foundation of China under Grant No. 08JJ0001; and
the construct program of the key discipline in hunan province.
\end{acknowledgments}

\end{document}